\documentclass [12pt]{article}
\usepackage {graphicx}

\title{Computation of current-voltage characteristics of weak links}

\author{ D.M. Gokhfeld }

\begin{document}

\maketitle

\begin{center}

\vskip0.5ex \textsl{ L.V. Kirensky Institute of Physics SD RAS,
Krasnoyarsk, 660036, Russia} \\ E-mail: gokhfeld@iph.krasn.ru \\

\end{center}

Simplified model for current-voltage characteristics of weak links
is suggested. It is based on approach which considers Andreev
reflections as responsible for the dissipative current through the
metallic Josephson junction. The model allows to calculate
current-voltage characteristics of weak links (superconductor -
normal metal - superconductor junctions, microbridges,
superconducting nanowires) for different thicknesses of the normal
layer at different temperatures. The current-voltage
characteristics of tin microbridges at different temperatures were
computed.

\section{Introduction}

Superconductor -- normal metal -- superconductor (SNS) junctions
have the current-voltage characteristics (CVCs) with the rich
peculiarities. Given certain parameters of junction, CVCs of SNS
junctions demonstrate the current peak at the small voltage, the
excess current, the subgarmonic gap structure and the negative
differential resistance at low bias voltage. Such nonlinear CVCs
make SNS junctions to be promising for application to low-noise
mixers in submillimetre-wave region \cite{nic3g,matsu}, switcher
\cite{mamal} or nanologic circuits \cite{hu4}.

Description of CVCs of SNS junctions was subject of many articles
and there were recognized key role of multiple Andreev reflections
at NS interfaces \cite{kbt,octfh,kgn,guza,brat,baav}.

The main features of CVCs enumerated above are successfully
described by K\"{u}mmel - Gunsenheimer - Nicolsky theory (KGN)
\cite{kgn} where time dependent Bogoliubov - de Gennes equations
are solved and wave packets of the nonequilibrium electrons and
holes are considered. KGN theory is applicable for relatively
thick and clean weak links where the normal metal layer N has the
thickness $2a$ larger than the coherence length of superconductor
and the inelastic mean free path $l$ larger than $2a$. Simplified
model in frame of KGN theory was developed by L.A.A. Pereira and
R. Nicolsky \cite{nic}. This model is relevant for the weak links
with thin superconducting banks S. The contribution of the
scattering states \cite{kgn} is omitted in \cite{nic}.

KGN and Pereira - Nicolsky model were applied earlier to describe
the experimental CVCs
\cite{nic2,PphC99,Pftt02,Pftt03,PphC04,go04}. Experience of the
applications demonstrates that oversimplified Pereira - Nicolsky
model gives only qualitative description. New simple modification
of KGN theory is proposed in this article. It is shown that CVCs
of SNS junctions can be computed without the all complex Ansatz of
KGN theory. I hope it will lead to more extensive using of the KGN
based approach to the calculation of weak link characteristics
than it was earlier
\cite{nic2,PphC99,hu4,Pftt02,Pftt03,PphC04,go04}.

\section{Current-voltage characteristics}

\subsection{Model}

Let us consider a voltage-biased SNS junction with a constant
electric field which is in negative z direction perpendicular to
the NS interfaces and exists in the N layer only (Fig. 1). The
normal layer has the thickness $2a$. The thickness of the
superconducting bank is $D-a >> 2a$.

\begin{figure}[htbp]
\centerline{\includegraphics[width=2.39in,height=2.26in]{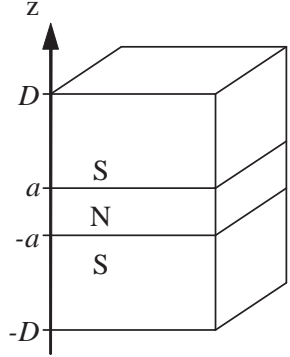}}
\caption{SNS junction modelled.} \label{fig1}
\end{figure}

Accordingly KGN \cite{kgn} the expression for CVC of SNS junction
with thick superconducting banks ($D-a >> 2a$) can be written as
following:

\begin{eqnarray}
\label{eq1}  I(V) = \frac{e\hbar}{2am^\ast } \sum\limits_{n =
1}^\infty {\exp \left( {
- \frac{2a}{l} n} \right)}\qquad\qquad\qquad\qquad\qquad\qquad\qquad\qquad\qquad\  \nonumber \\
\int\limits_{ - \Delta + neV}^{\Delta + eV} dE \: \sum\limits_{r}
g_r \left( E \right) P_N \left( E \right) k_{zF} \tanh \left(
{\frac{E}{k_b T}} \right) \qquad\quad \nonumber \\ + \frac{V}{R_N}
\,,\quad
\end{eqnarray}

\noindent where $m^\ast$ is the effective mass of electron,
$g_r(E)$ is the two dimensional density of states, $P_{N}(E)$ is
the probability of finding of the quasiparticles with the energy
$E$ in the N region of the thickness $2a$, $l$ is the inelastic
mean free path and $R_{N}$ is the resistance of the N region,
$\Delta$ is the value of energy gap of superconductor at the
temperature $T$, $k_{zF}$ is the z component of Fermi wave vector
of quasiparticles, $n$ is the number of Andreev reflections which
quasiparticles undergo before they move out of the pair potential
well.

Eq.(\ref{eq1}) is for the time averaged current that includes the
voltage dependence only within the integral limits.

\subsection{Density of states}

To operate on Eq.(\ref{eq1}) one should calculate the density of
states \cite{plegk}:

\begin{equation}
\label{eq1a} g_r(E) = \frac{A}{\pi} \sum\limits_{r} k_{zF,r}
\left| {\frac{dE}{dk_{zF}}} \right|^{-1}_{k_{zF,r}},
\end{equation}
\noindent where $A$ is the normal layer area, $k_{zF,r}$ defines
the value of $k_{zF}$ for which $E_r = E$.

The energy spectrum $E_r(k_{zF})$ consists of the spatially
quantized bound states and the quasicontinuum scattering states.
The energy eigenvalue equation for the spatially quantized bound
Andreev states \cite{kgn} is transcendental and calculated
numerically only:

\begin{equation}
\label{eq2er} E_r \left( {k_{zF} } \right) = \frac{\hbar ^2k_{zF}
}{2am^\ast}\left( {r\pi + \arccos \frac{E_r }{\Delta }} \right),
\end{equation}

\noindent where $r$ = 0,1,2,\ldots

Let us simplify Eq.(\ref{eq2er}). The expansion of
arccos($E_{r}/\Delta )$ in (\ref{eq2er}) to Taylor series $(\pi/2
- E_{r}/\Delta $ +\ldots ) up to second term and subsequent
expressing of $E_r(k_{zF})$ are executed:

\begin{equation}
\label{eq3c1} E_r \left( {k_{zF} } \right) \approx \frac{\hbar
^2k_{zF} }{2am^\ast}\pi \left( {r + \textstyle{1 \over 2}} \right)
/ \left( {1 + \\ \frac{\hbar ^2k_{zF} }{2am^\ast \Delta}} \right)
\end{equation}

\begin{figure}[htbp]
\centerline{\includegraphics[width=87.5mm,height=63mm]{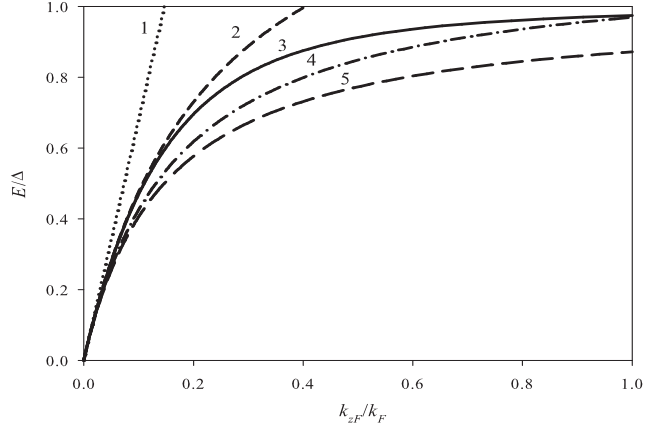}}
\caption{Energy of the bound Andreev state with $r=0$; $2a=5000$
{\AA}; $\Delta = 0.57$ meV; $k_F = 1.62$ {\AA}$^{-1}$. 1)
Eq.(\ref{eq4c}), $C=0$; 2) Eq.(\ref{eq4c}), $C=1$; 3) the exact
solution of Eq.(\ref{eq2er}); 4) Eq.(\ref{eq4c}), $C= \pi/2(1-
am^\ast\Delta/\hbar^2k_{F})$; 5) Eq.(\ref{eq4c}), $C=\pi/2$.}
\label{fig2st}
\end{figure}

Dependence $E_r(k_{zF})$ (\ref{eq3c1}) (curve 2) and the numerical
solution of Eq.(\ref{eq2er}) (curve 3) are shown in Fig. 2. The
better agreement with the numerical solution of Eq.(\ref{eq2er})
is attained by insertion of the correcting multiplier $C$ before
$\hbar ^2k_{zF} / 2am^\ast \Delta$ in Eq.(\ref{eq3c1}):

\begin{equation}
\label{eq4c} E_r \left( {k_{zF} } \right) \approx \frac{\hbar
^2k_{zF} }{2am^\ast}\pi \left( {r + \textstyle{1 \over 2}} \right)
/ \left( {1 + C\frac{\hbar ^2k_{zF} }{2am^\ast \Delta}} \right)
\end{equation}

If $C=0$ then the spectrum of Pereira - Nicolsky model is
reproduced (curve 1, Fig. 2). R. K\"{u}mmel used Eq.(\ref{eq4c})
with $C=\pi/2$ \cite{kue_pr} for approximated calculation of the
energy spectrum (curve 5, Fig. 2). I suggest the variable
multiplier $C= \pi/2(1- am^\ast\Delta/\hbar ^2k_{F})$ for $C>1$
and $C=1$ otherwise. Such choice of $C$ provides a good agreement
of Eq.(\ref{eq4c}) (curve 4, Fig. 2) with the numerical solution
of Eq.(\ref{eq2er}) for different relation of
$a,m^\ast,\Delta,k_F$.

The density of the bound states with (\ref{eq4c}) becomes

\begin{equation}
\label{eq5gbs} g_{r}\left( E \right) = \frac{A}{\pi }\left(
{\frac{2m^\ast a}{\hbar ^2}} \right)^2\sum\limits_r {\frac{E}{\pi
^2\left( {r + \textstyle{1 \over 2}} \right)^2\left( {1 -
C\frac{E}{\pi \left( {r +\textstyle{1 \over 2}} \right)\Delta }}
\right)^3}}
\end{equation}

For quasiparticles from the quasicontinuum states the energy
spectrum is approximated by the continuous BCS spectrum of a
homogeneous superconductor \cite{kgn,plegk}:

\begin{equation}
\label{eq_ess} E(k_{zF}) = \sqrt{\left(\frac{\hbar^2}{2 m^\ast}
\left(k_F^2-k_{zF}^2\right)\right)^2 + {\Delta^*}^2}
\end{equation}

\begin{figure}[htbp]
\centerline{\includegraphics[width=87.5mm,height=63mm]{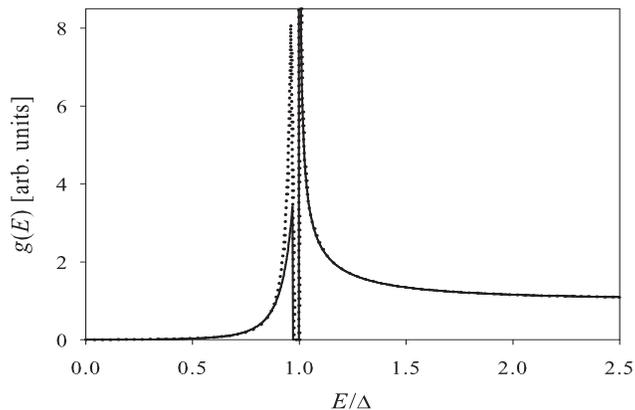}}
\caption{Density of states $g(E)$ of an SNS junction with thick
superconducting banks resulted by \cite{plegk} (dotted line) and
 $g(E)$ calculated by Eq.(\ref{eq4c}) with $C= \pi/2(1- am^\ast\Delta/\hbar ^2k_{F})$ (solid line). $D=70000$ {\AA};
$2a=5000$ {\AA}; $T_c=3.77$ K; $k_F = 1.62$ {\AA}$^{-1}$.}
\label{fig3dos}
\end{figure}

In the case of SNS junction with thick superconducting banks the
effective energy gap $\Delta^*$ equals $\Delta$. The density of
the quasicontinuum scattering states is

\begin{equation}
\label{eq7gss} g(E) = \frac{A}{\pi^2 } \frac{2m^\ast}{\hbar ^2}
k_F D \frac{E}{\sqrt {E^2 - \Delta ^2} }
\end{equation}

The density of states resulted is shown in Fig. 3.

\subsection{Current density}

The probability $P_{N}(E)$ of finding of the quasiparticles with
the energy $E$ in the N region is given by Eq.(2.19) of
\cite{kgn}:

\begin{equation}
\label{eq_pn} P_N(E) = \frac{2a}{2 a + 2 \lambda}
\end{equation}
\noindent with the penetration depth  $ \lambda =
\frac{\hbar^2}{m^\ast} \frac{k_{zF}}{\sqrt{\Delta ^2 - E^2}}$ for
$E < \Delta$, $\lambda < D - a$ and $\lambda = D - a$ otherwise.

For the quasiparticles from the scattering states $P_N(E)= 2a/2D$.
Let us accept for the sake of simplicity $\lambda >> a$ and
therefore $P_N(E)= 2a/2\lambda$ for the bound states.

The current density of quasiparticles from the bound states is
resulted with (\ref{eq5gbs}):

\begin{eqnarray}
\label{eq_jbs} j_{bs} (V) = \frac{e{m^\ast}^2 a^2}{2 \pi^3
\hbar^5} \sum\limits_n \exp \left( { - \frac{2a}{l} n} \right)
\qquad\qquad\qquad\qquad\qquad\qquad\qquad \nonumber
\\* \int\limits_{ - \Delta + neV}^\Delta {dE\sum\limits_r
{\frac{\left| E \right|\sqrt {\Delta ^2 - E^2}}{\left( {r +
\textstyle{1 \over 2}} \right)\left( {1 - C\frac{\left| E
\right|}{\pi \Delta \left( {r +\textstyle{1 \over 2}} \right)}}
\right)^3}} \tanh \left( \frac{E}{2k_B T} \right) }
\end{eqnarray}

With using (\ref{eq7gss}) the current density of quasiparticles
from the quasicontinuum states is resulted:

\begin{eqnarray}
\label{eq8} \nonumber j_{ss} (V) = \frac{e}{4 \pi^2 \hbar } k_F
\sum\limits_n \exp \left( { - \frac{2a}{l} n} \right)
\qquad\qquad\qquad\qquad\qquad\qquad\qquad \\* \int\limits_{E_1
}^{\Delta + eV} dE\frac{E \sqrt {k_F ^2 - \frac{2m^\ast}{\hbar
^2}\sqrt {E^2 - \Delta ^2} }}{\sqrt {E^2 - \Delta ^2} } \tanh
\left( \frac{E}{2k_B T} \right)
\end{eqnarray}

Here and further $E_{1}$ = $-\Delta + neV$ for $-\Delta + neV \ge
\Delta $ and $E_{1}=\Delta $ otherwise.

Neglecting the small term I have

\begin{equation}
\label{eq9} j_{ss} (V) = \frac{e}{4 \pi^2 \hbar } {k_F}^2
\sum\limits_n {\exp \left( { - \frac{2a}{l} n}
\right)\int\limits_{E_1 }^{\Delta + eV} {dE\frac{E\tanh \left( {E
/ 2k_B T} \right)}{\sqrt {E^2 - \Delta ^2} }} }
\end{equation}

If $eV >> k_{B}T$, $\Delta $ the integral in (\ref{eq9}) can be
transformed and the excess current density is resulted:

\begin{equation}
\label{eq10} j_{ex} (V) = \frac{e}{2 \pi^2 \hbar } {k_F}^2 \Delta
\tanh \left( \frac{eV}{2k_B T} \right)\exp \left( - \frac{2a}{l}
\right)
\end{equation}

This excess current density is the same as one in KGN (Eq.(4.12)
in \cite{kgn}).

Note that $j_{bs}(V)$ dependence does not change practically if
the second summation in (\ref{eq_jbs}) is interrupted at $r = 0$.
Therefore the total current density is

\begin{eqnarray}
\label{eq11tot} j(V) = \sum\limits_n \exp \left( { - \frac{2a}{l}
n} \right) \Biggl \lbrace \frac{2 e {m^\ast}^2 a^2}{\pi^3 \hbar^5}
\int\limits_{-\Delta + neV}^\Delta dE {\frac{\left| E \right|\sqrt
{\Delta ^2 - E^2}}{\left( 1 - C \frac{2 \left| E \right|}{\pi
\Delta} \right)^3}} \tanh{\left( \frac{E}{2k_B T} \right)}
\nonumber
\\* + \frac{e {k_F}^2 }{4 \pi^2 \hbar} \int\limits_{E_1}^{\Delta + eV}
dE \frac{E}{\sqrt {E^2 - \Delta^2}} \tanh{\left( \frac{E}{2k_B T}
\right)} \Biggr \rbrace + \frac{V}{R_{N} A} \qquad\qquad
\end{eqnarray}
$C= \pi/2(1- am^\ast\Delta/\hbar ^2k_{F})$ for $C>1$ and $C=1$
otherwise; $E_{1}$ = $-\Delta + neV$ for $-\Delta + neV \ge \Delta
$ and $E_{1}=\Delta $ otherwise.

Eq.(\ref{eq11tot}) is the main result of this simplified model.
The model allows to calculate CVCs of weak links for different
thicknesses of the normal layer at different temperatures. The
subgarmonic gap structure, the excess current and the current peak
at the small voltage are reproduced on CVCs.

\subsection{Comparison with experimental current-voltage characteristics}

The model was used to compute two sets of CVCs of tin
microbridges. These detailed measurements of the current biased
CVCs were performed by V.N. Gubankov, V.P. Kosheletz, G.A.
Ovsyannikov in 1977-1981 \cite{guba,guba1} (Fig. 4) and M.
Octavio, W.J. Skocpol, M. Tinkham in 1978 \cite{oct} (Fig. 5).
Both sets of CVCs have the similar peculiarities: the subgarmonic
gap structure, the current peak at the small voltage and the
excess current.

\begin{figure}[htbp]
\centerline{\includegraphics[width=87.5mm,height=63mm]{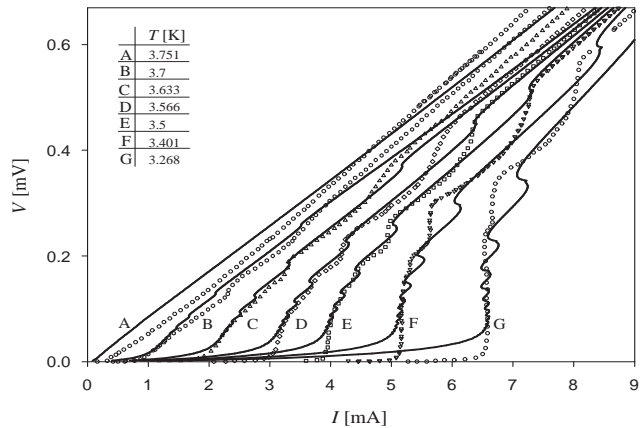}}
   \caption{The current-voltage characteristic of Sn microbridges. Experiment \cite{guba1} (points) and calculations (solid lines).}
\label{figcvcg}
\end{figure}

Comparison of the computed $I(V)$ curves and the experimental $V$
vs. $I$ dependencies displays satisfactory agreement at
temperatures smaller than $0.99 T_c$. Presented in Fig. 4 and Fig.
5 curves were calculated with the reasonable parameters: the
critical temperature, the energy gap at zero temperature, the
Fermi wave vector of Sn ($T_c = 3.77$ K, $\Delta_0 = 0.57$ meV,
$k_F = 1.62$ {\AA}$^{-1}$). The length of microbridges $2a$ is
5000 {\AA} and $l=15 a$. The BCS dependence of $\Delta$ on $T$ was
used.

\begin{figure}[htbp]
\centerline{\includegraphics[width=87.5mm,height=63mm]{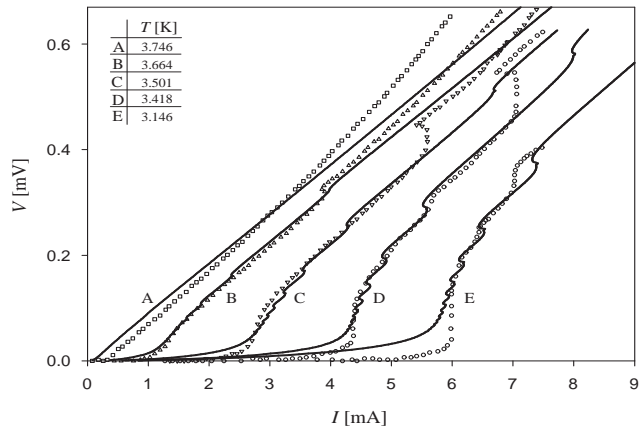}}
   \caption{The current-voltage characteristic of Sn microbridges.  Experiment \cite{oct} (points) and calculations (solid lines).}
\label{figcvco}
\end{figure}

The high voltages regions on experimental CVCs are close by the
computed curves at higher temperatures. It is possibly reasoned by
selfheating occurred at high voltages in these experiments
\cite{guba1,oct}. Some discrepancy of the computed curves and the
experimental points at low voltages is because there were the
current-biased CVCs in experiments instead voltage-biased one.
Agreement of the model and the experimental CVCs disappears at
temperatures near $T_c$: the calculated current peak and the
excess current is smaller than corresponding experimental currents
(e.g. CVCs at 3.751 K in Fig. 4 and at 3.746 K in Fig.5).

\section{Conclusion}

Simplified model for calculation of current-voltage
characteristics of the weak links (SNS junctions, microbridges,
superconducting nanowires) was developed. This model makes the KGN
approach \cite{kgn} more convenient for description of
experiments. The model was applied for computation of the
current-voltage characteristics of tin microbridges at different
temperatures.\\

\section*{Acknowledgements}

I am thankful to D.A. Balaev, R. K\"{u}mmel and M.I. Petrov for
fruitful discussions. This work is supported by program of
President of Russian Federation for support of young scientists
(grant MK 7414.2006.2), Krasnoyarsk Regional Scientific Foundation
(grant 16G065), program of presidium of Russian academy of science
"Quantum macrophysics" 3.4, Lavrent'ev competition of young
scientist projects (project 52).


\begin{thebibliography}{30}

\bibitem{nic3g} Y.A. Gorelov, L.A.A. Pereira, A.M. Luiz, R. Nicolsky. Physica C
282-287, 2491 (1997).
\bibitem{matsu} T. Matsui, H. Ohta. Supercond. Sci. Technol. 12, 859 (1999).
\bibitem{mamal} A.G. Mamalis, D.M. Gokhfeld, S.V. Militsyn, M.I. Petrov,
D.A. Balaev, K.A. Shaihutdinov, S.G. Ovchinnikov, V.I. Kirko, I.N.
Vottea. Journ. of Materials Processing Technology 161, 42 (2005).
\bibitem{hu4} C.H. Hu, J.F. Jiang, Q.Y. Cai. Supercond. Sci. Technol. 15, 330 (2002).
\bibitem{kbt} T.M. Klapwijk, G.E. Blonder, M. Tinkham. Physica B 109\&110, 1657 (1982).
\bibitem{octfh} K. Flensberg, J. Bindslev Hansen, M. Octavio. Phys. Rev. B 38, 8707 (1988).
\bibitem{kgn} R. K\"{u}mmel, U. Gunsenheimer, R. Nicolsky. Phys. Rev. B 42, 3992 (1990).
\bibitem{guza} U. Gunsenheimer, A.D. Zaikin. Phys. Rev. B 50, 6317 (1994).
\bibitem{brat} E.N. Bratus', V.S. Shumeiko, E.V. Bezuglyi, G. Wendin. Phys.
Rev. B 55, 12666 (1997).
\bibitem{baav} A. Bardas, D. Averin. Phys. Rev. B 56, 8518 (1997).
\bibitem{nic} L.A.A. Pereira, R. Nicolsky. Physica C 282-287, 2411 (1997).
\bibitem{nic2} L.A.A. Pereira, A.M. Luiz, R. Nicolsky. Physica C 282-287, 1529 (1997).
\bibitem{PphC99} M.I. Petrov, D.A. Balaev, D.M. Gohfeld, S.V. Ospishchev, K.A. Shaihutdinov, K.S. Aleksandrov. Physica C 314, 51 (1999).
\bibitem{Pftt02}  M.I. Petrov, D.A. Balaev, D.M. Gokhfeld, K.A. Shaikhutdinov, K.S. Aleksandrov. Fiz. Tverd. Tela 44, 1179 (2002) [Phys. Solid State 44, 1229 (2002)].
\bibitem{Pftt03} M.I. Petrov, D.A. Balaev, D.M. Gokhfeld, K.A. Shaikhutdinov. Fiz. Tverd. Tela 45, 1164 (2003) [Phys. Solid State 45, 1219 (2003)].
\bibitem{PphC04} M.I. Petrov, D.M. Gokhfeld, D.A. Balaev, K.A. Shaihutdinov, R. K\"{u}mmel. Physica C 408, 620 (2004).
\bibitem{go04} D.M. Gokhfeld, D.A. Balaev, K.A. Shaykhutdinov, S.I. Popkov, M.I. Petrov. cond-mat/0410112 (2004).
\bibitem{plegk} H.Plehn, U. Gunsenheimer, R. K\"{u}mmel. Journ. Low Temp. Phys. 83, 71 (1991).
\bibitem{kue_pr} R. K\"{u}mmel. Private communications.
\bibitem{guba} V.N. Gubankov, V.P. Kosheletz, G.A. Ovsyannikov. Journ. Exp. i Teoretich. Fiziki 73, 1435 (1977).
\bibitem{guba1} V.N. Gubankov, V.P. Kosheletz, G.A. Ovsyannikov. Fizika Nizkikh Temperatur 7, 277 (1981).
\bibitem{oct} M. Octavio, W.J. Skocpol, M. Tinkham. Phys. Rev. B 17, 159 (1978).

\end{thebibliography}
\end{document}